\documentclass{llncs}

\usepackage{graphicx}
\usepackage{textcomp}
\usepackage{array,ragged2e}
\usepackage{xcolor}
\usepackage{colortbl}
\usepackage{makecell}
\usepackage{pgfplots}
\usepackage{amsmath}
\usepackage{multicol}
\usepackage{tabto}
\usepackage{wrapfig}
\usepackage{amssymb}
\usepackage{float}

\begin{document}

\mainmatter              
\title{Designing and Implementing Data Warehouse \\for Agricultural Big Data}

\vspace{-5mm}
\titlerunning{Hamiltonian Mechanics}  
%
\vspace{-3mm}

\author{Vuong M. Ngo \and Nhien-An Le-Khac \and M-Tahar Kechadi}

\institute{School of Computer Science, University College Dublin,\\
		Belfield, Dublin 4, Ireland\\
\email{vuong.ngo@ucd.ie, an.lekhac@ucd.ie, tahar.kechadi@ucd.ie}
}

\maketitle
\vspace{-3mm}
\begin{abstract}
	In recent years, precision agriculture that uses modern information and communication technologies is becoming very popular. Raw and semi-processed agricultural data are usually collected through various sources, such as: Internet of Thing (IoT), sensors, satellites, weather stations, robots, farm equipment, farmers and agribusinesses, etc. Besides, agricultural datasets are very large, complex, unstructured, heterogeneous, non-standardized, and inconsistent. Hence, the agricultural data mining is considered as Big Data application in terms of volume, variety, velocity and veracity. It is a key foundation to establishing a crop intelligence platform, which will enable resource efficient agronomy decision making and recommendations. In this paper, we designed and implemented a continental level agricultural data warehouse by combining Hive, MongoDB and Cassandra. Our data warehouse capabilities: (1) flexible schema; (2) data integration from real agricultural multi datasets; (3) data science and business intelligent support; (4) high performance; (5) high storage; (6) security; (7) governance and monitoring; (8) replication and recovery; (9) consistency, availability and partition tolerant; (10) distributed and cloud deployment. We also evaluate the performance of our data warehouse.

\end{abstract}

\vspace{-3mm}
\section{Introduction}
\vspace{-2mm}

In 2017 and 2018, annual world cereal productions were 2,608 million tons \cite{USDA.2018} and 2,595 million tons \cite{FAO-CSDB.2018}, respectively. However, there were also around 124 million people in 51 countries faced food crisis and food insecurity \cite{FAO-FSIN.2018}. According to United Nations \cite{UnitedNations.2017}, we need an increase 60\% of cereal production to meet 9.8 billion people needs by 2050. To satisfy the massively increase demand for food, crop yields must be significantly increased by using new farming approaches, such as precision agriculture. As reported in \cite{Eurobarometer.2016}, precision agriculture is vitally important for the future and can make a significant contribution to food security and safety. Besides, precision agriculture promises both high quantity and quality of its products with minimum of resource usage, such as water, energy, fertilisers, and pesticides \cite{Schrijver.2016}.

The precision agriculture's current mission is to use the decision-support system based on Big Data approaches to provide precise information for more control of farming efficiency and waste, such as awareness, understanding, advice, early warning, forecasting and financial services. An efficient agricultural data warehouse (DW) is required to extract useful knowledge and support decision-making. However, currently there are very few reports in the literature that focus on the design of efficient DWs with the view to enable Agricultural Big Data analysis and mining. The design of large scale agricultural DWs is very challenging. Moreover, the precision agriculture system can be used by different kinds of users at the same time, for instance by both farmers and agronomists. Every type of user needs to analyse different information sets thus requiring specific analytics. The agricultural data has all the features of Big Data: 
\begin{enumerate}
\item Volume: The amount of agricultural data is rapidly increasing and is intensively produced by endogenous and exogenous sources. The endogenous data is collected from operation systems, experimental results, sensors, weather stations, satellites and farm equipment. The systems and devices in the agricultural ecosystem can connect through IoT. The exogenous data concerns the external sources, such as farmers, government agencies, retail agronomists and seed companies. They can help with information about local pest and disease outbreak tracking, crop monitoring, market accessing, food security, products, prices and knowledge.

\item Variety: Agricultural data has many different forms and formats, such as structured and unstructured data, video, imagery, chart, metrics, geo-spatial, multi-media, model, equation and text.

\item Velocity: The produced and collected data increases at high rate, as sensing technologies and other mobile devices are becoming more efficient and cheaper. The datasets must be cleaned, aggregated and harmonised in real-time.

\item Veracity: The tendency of agronomic data is uncertain, inconsistent, ambiguous and error prone because the data is gathered from heterogeneous sources, sensors and manual processes.
\end{enumerate}

In this research, firstly, we analyze popular DWs to handle agricultural Big Data. Secondly, an agricultural DW is designed and implemented by combining Hive, MongoDB, Cassandra, and constellation schema on real agricultural datasets. Our DW has enough main features of a DW for agricultural Big Data. These are: (1) high storage, high performance and cloud computing adapt for the volume and velocity features; (2) flexible schema and integrated storage structure to adapt the variety feature; (3) data ingestion, pre-processing, governance, monitoring and security adapt for the veracity feature. Thirdly, the effective business intelligent support is illustrated by executing complex HQL/SQL queries to answer difficult data analysis requests. Besides, an experimental evaluation is conducted to present good performance of our DW storage. The rest of this paper is organised as follows: in the next Section, we reviewed the related work. In Sections 3, 4, and 5, we presented solutions for the above goals, respectively. Finally, Section 6 gives some concluding remarks.

\section{Related Work}
\vspace{-1mm}
Data mining can be used to design an analysis process for exploiting big agricultural datasets. Recently, many papers have been published that exploit machine learning algorithms on sensor data and build models to improve agricultural economics, such as \cite{Pantazi.2016}, \cite{Park.2016}, \cite{Feng.2017} and \cite{Rupnik.2018}. In these, the paper \cite{Pantazi.2016} predicted crop yield by using self-organizing-maps supervised learning models; namely supervised Kohonen networks, counter-propagation artificial networks and XY-fusion. The paper \cite{Park.2016} predicted drought conditions by using three rule-based machine learning; namely random forest, boosted regression trees, and Cubist. The paper \cite{Feng.2017} estimated water evapotranspiration by using extreme learning machine and generalized regression neural network models which based on daily temperature data in six meteorological stations. Finally, the paper \cite{Rupnik.2018} predicted pest population dynamics by using time series clustering and structural change detection which detected groups of different pest species. However, the proposed solutions are not satisfied the problems of agricultural Big Data, such as data integration, data schema, storage capacity, security and performance.

From a Big Data point of view, the papers  \cite{Kamilaris.2018} and \cite{Schnase.2017} have proposed “smart agricultural frameworks”.  In \cite{Kamilaris.2018}, the platform used Hive to store and analyse sensor data about land, water and biodiversity which can help increase food production with lower environmental impact. In \cite{Schnase.2017}, the authors moved toward a notion of climate analytics-as-a-service by building a high-performance analytics and scalable data management platform which is based on modern infrastructures, such as Amazon web services, Hadoop and Cloudera. However, the two papers did not discuss how to build and implement a DW for a precision agriculture.

Our approach is inspired by papers \cite{Schulze.2007}, \cite{Schuetz.2018}, \cite{Nilakanta.2008} and \cite{Ngo.2018} which presented ways of building a DW for agricultural data. In \cite{Schulze.2007}, the authors extended entity-relationship model for modelling operational and analytical data which is called the multi-dimensional entity-relationship model. They introduced new representation elements and showed the extension of an analytical schema. In \cite{Schuetz.2018}, a relational database and an RDF triple store, were proposed to model the overall datasets. In that, the data are loaded into the DW in RDF format, and cached in the RDF triple store before being transformed into relational format. The actual data used for analysis was contained in the relational database. However, as the schemas in \cite{Schulze.2007} and \cite{Schuetz.2018} were based on entity-relationship models, they cannot deal with high-performance, which is the key feature of a data warehouse.

In \cite{Nilakanta.2008}, a star schema model was used. All data marts created by the star schemas are connected via some common dimension tables. However, a star schema is not enough to present complex agricultural information and it is difficult to create new data marts for data analytics. The number of dimensions of DW proposed by \cite{Nilakanta.2008} is very small; only 3-dimensions – namely, Species, Location, and Time. Moreover, the DW concerns livestock farming.  Overcoming disadvantages of the star schema, the paper \cite{Ngo.2018} proposed a constellation schema for an agricultural DW architecture in order to facilitate quality criteria of a DW. However, it does not describe how to implement the proposed DW.

 \vspace{-3mm}
\section{Analyzing Cassandra, MongoDB and Hive in agricultural Big Data}
\vspace{-1mm}
	In general, a DW is a federated repository for all the data that an enterprise can collect through multiple heterogeneous data sources belonging to various enterprise's business systems or external inputs \cite{Golfarelli-Rizzi.2009}, \cite{Inmon.2005}. A quality DW should adapt many important criteria \cite{Adelman-Moss.2000}, \cite{Kimball-Ross.2013}, such as: (1) Making information easily accessible; (2) Presenting and providing right information at the right time; (3) Integrating data and adapting to change; (4) Achieving tangible and intangible benefits; (5) Being a secure bastion that protects the information assets; and (6) Being accepted by DW users. So, to build an efficient agricultural DW, we need to take into account these criteria.
	
	Currently, there are many popular databases that support efficient DWs, such as such as Redshift, Mesa, Cassandra, MongoDB and Hive. Hence, we are analyzing the most popular and see which is the best suited for our data problem. In these databases, Redshift is a fully managed, petabyte-scale DW service in the cloud which is part of the larger cloud-computing platform Amazon Web Services \cite{Amazon.2018}. Mesa is highly scalable, petabyte data warehousing system which is designed to satisfy a complex and challenging set of users and systems requirements related to Google’s Internet advertising business \cite{Gupta.2016}. However, Redshift and Mesa are not open source. While, Cassandra, MongoDB and Hive are open source databases, we want to use them to implement agriculture DW. Henceforth, the Cassandra and MongoDB terms are used to refer to DWs of Cassandra and MongoDB databases.
	
	There are many papers studying Cassandra, MongoDB and Hive in the view of general DWs. In the following two subsections, we present advantages, disadvantages, similarities and differences between Cassandra, MongoDB and Hive in the context of agricultural DW. Specially, we analyze to find how to combine these DWs together to build a DW for agricultural Big Data, not necessarily best DW.
		
	\vspace{-2mm}
	\subsection{Advantages and disadvantages}
	\vspace{-2mm}
	  	Cassandra, MongoDB and Hive are used widely for enterprise DWs. Cassandra\footnote{http://cassandra.apache.org} is a distributed, wide-column oriented DW from Apache that is highly scalable and designed to handle very large amounts of structured data. It provides high availability with no single point of failure, tuneable and consistent. Cassandra offers robust support for transactions and flexible data storage based on ideas of DynamoDB and BigTable \cite{Hewitt.2016}, \cite{Neeraj.2015}. While, MongoDB\footnote{http://mongodb.com} is a powerful, cross-platform, document oriented DW that provides, high performance, high availability, and scalability \cite{Chodorow.2013}, \cite{Hows.2015}. It works on concept of collection and document, JSON-like documents, with dynamic schemas. So, documents and data structure can be changed over time. Secondly, MongoDB combines the ability to scale out with features, such as ad-hoc query, full-text search and secondary index. This provides powerful ways to access and analyze datasets. Hive\footnote{http://hive.apache.org} is an SQL data warehouse infrastructure on top of Hadoop\footnote{http://hadoop.apache.org} for writing and running distributed applications to summarize Big Data \cite{Du.2018}, \cite{Lam.2016}. Hive can be used as an online analytical processing (OLAP) system and provides tools to enable data extract - transform - load (ETL). Hive's metadata structure provides a high-level, table-like structure on top of HDFS (Hadoop Distributed File System). That will significantly reduce the time to perform semantic checks during the query execution. Moreover, by using Hive Query Language (HQL), similar to SQL, users can make simple queries and analyse the data easily.
		
		Although, the three DWs have many advantages and have been used widely, they have major limitations. These limitations impact heavily on their use as agricultural DW.
		\begin{enumerate}
		\item In Cassandra: (1) Query Language (CQL) does not support joint and subquery, and has limited support for aggregations that are difficult to analyze data; (2) Ordering is done per-partition and specified at table creation time. The sorting of thousands or millions of rows can be fast in development but sorting billion ones is a bad idea; (3) A single column value is recommended not be larger than 1MB that is difficult to contain videos or high quality images, such as LiDAR images, 3-D images and satellite images.
		\item In MongoDB: (1) The maximum BSON document size is 16MB that is difficult to contain large data such as video, audio and high quality image; (2) JSON’s expressive capabilities are limited because the only types are null, boolean, numeric, string, array, and object; (3) We cannot automatically rollback more than 300 MB of data. If we have more than that, manual intervention is needed.
		\item Hive is not designed for: (1) Online transaction processing; (2) Real-time queries; (3) Large data on network; (4) Trivial operations; (5) Row-level update; and (6) Iterative execution.
		\end{enumerate}

	\vspace{-5mm}
\subsection{Feature Comparison}
	\vspace{-1mm}

Table \ref{tab1} lists technical features used to compare Hive, MongoDB and Cassandra. For the ten overview features given in section A of Table \ref{tab1}, the three DWs differ in data schema, query language and access methods. However, they all support map reduce. Moreover, the ETL feature is supported by Hive, limited to Cassandra and unsupported by MongoDB. The full-text search feature is only supported by MongoDB. The secondary index and ad-hoc query features are supported by Hive and MongoDB but not or restricted by Cassandra. The 9\textsuperscript{th} feature being the Consistency – Availability – Partition tolerant classification (CAP) theorem says how the database system behaves when facing network instability. It implies that in the presence of a network partition, one has to choose between consistency and availability. Hive and Cassandra choose availability. While, MongoDB chooses consistency. Finally, the structure of Hive and MongoDB are master - slave while Cassandra has peer - to - peer structure.
		
		\begin{table}
		\vspace{-2mm}
		\caption{Technical Features}
		\vspace{-4mm}
		\begin{center}
		\scriptsize
		\begin{tabular}{|c|c|c|c|c|}
		\hline
		\textbf{No.} & \textbf{Features} & \textbf{Hive}& \textbf{MongoDB}& \textbf{Cassandra} \\[1pt]
		\hline
		\multicolumn{5}{|c|}{\textbf{\textit{A. Overview Features}}}\\[1pt]
		\hline
		1 & Data scheme & Yes & No-Schema &	Flexible Schema\\
		\hline
		2 & Query language & HQL & JS-like syntax & CQL\\
		\hline
		3 & Accessing method & JDBC, ODBC, Thrift & JSON & Thrift \\
		\hline
		4 & Map reduce  & Yes  & Yes & Yes \\
		\hline
		5 & ETL & Yes & No  & Limited \\
		\hline
		6 & Full-text search & No & Yes & No \\	
		\hline
		7 & Ad-hoc query & Yes & Yes & No\\	
		\hline
		8 & Secondary index & Yes & Yes & Restricted\\
		\hline
		9 & CAP &  AP & CP & AP  \\
		\hline
		10 & Structure & Master – Slave & Master – Slave & Peer – to – Peer\\
		\hline
		\multicolumn{5}{|c|}{\textbf{\textit{B. Industrial Features}}}\\[1pt]
		\hline
		1 & Governance & Yes (via Hadoop) & Yes & Yes (via JME) \\
		\hline
		2 & Monitoring & Yes & Yes & Yes \\
		\hline
		3 & Data Lifecycle Management & Yes (via Hadoop) & Yes & Yes \\
		\hline
		4 & Workload Management & Yes & Yes & Yes \\
		\hline
		5 & Replication-Recovery & Yes & Yes & Yes\\
		\hline
		\end{tabular}
		\label{tab1}
		\end{center}
		\vspace{-6mm}
		\end{table}

		The section B of Table \ref{tab1} describes five industrial features, such as governance, monitoring, data lifecycle management, workload management, and replication-recovery. All of Hive, MongoDB and Cassandra support these features. Hive supports governance and data lifecycle management features via Hadoop. Cassandra is based on Java Management Extensions (JME) for governance.

		\begin{table}
		\vspace{-4mm}
		\caption{Data Management and Data Warehouse Features}
		\vspace{-5mm}
		\begin{center}
		\scriptsize
		\begin{tabular}{|c|c|c|c|c|}
		\hline
		\textbf{No.} & \textbf{Features} & \textbf{Hive}& \textbf{MongoDB}& \textbf{Cassandra} \\[1pt]
		\hline
		\multicolumn{5}{|c|}{\textbf{\textit{A. Data Management Features}}}\\[1pt]
		\hline
		1 & Security & Yes & Yes & Yes \\
		\hline
		2 & High Storage Capacity & Yes (best) & Yes & Yes \\
		\hline
		3 & Data Ingestion and Pre-processing & Yes & Yes & No \\
		\hline
		\multicolumn{5}{|c|}{\textbf{\textit{B. Data Warehouse Features}}}\\[1pt]
		\hline
		1 & Business Intelligent & Very good & Limited & Good \\
		\hline
		2 & Data Science & Very good & Limited & Limited \\
		\hline
		3 & High Performance & Non-real time & Real time & Real time \\
		\hline
		\end{tabular}
		\label{tab2}
		\end{center}
		\vspace{-5mm}
		\end{table}

	The data management and DW features are described in section A and section B of Table \ref{tab2}, respectively. The data management features are security, high storage capacity, and data ingestion and pre-processing. The DWs have support for these features, except Cassandra does not support for data ingestion and pre-processing. Hive has the best for high storage capacity. The DW features are business intelligent, data science and high performance. Hive supports well business intelligent and data science but it is not suitable for real-time performance. MongoDB is very fast but it is limited in supporting for business intelligent and data science. Cassandra also is very fast and supports business intelligent but has limited capabilities for data science.

\vspace{-3mm}
\section{Agricultural Data Warehouse}
	
	\vspace{-2mm}
	The general architecture of a typical DW includes four separate and distinct modules being Raw Data, ETL, Integrated Information and Data Mining. In the scope of this paper, we focus on the Integrated Information module which is a logically a centralised repository. It includes DW storage, data marts, data cubes and OLAP engine.

	The DW storage is organised, stored and accessed using a suitable schema defined in the metadata. It can be either directly accessed or used to creating data marts which is usually oriented to a particular business function or enterprise department. A data cube is a data structure that allows fast analysis of data according to the multiple dimensions that define a business problem. The data cubes are created by the OLAP engine.
	
	\vspace{-3mm}
	\subsection{OLAP}
	\vspace{-2mm}
	OLAP is a category of software technology that provides the insight and understanding of data in multiple dimensions through fast, consistent, interactive access to enable analysts or managers to make better decisions. By using roll-up, drill-down, slice-dice and pivot operations, OLAP performs multidimensional analysis in a wide variety of possible views of information that provide complex calculations, trend analysis and sophisticated data modelling with a short execution time. So, OLAP is a key way to exploit information in a DW to allow end-users to analyze and explore data in multidimensional views.
	
	The OLAP systems are categorised into three types: namely relational OLAP (ROLAP), multidimensional OLAP (MOLAP) and hybrid OLAP (HOLAP):
	\begin{enumerate}
	\vspace{-1mm}
	\item ROLAP uses relational or extended-relational database management system to store and manage DW. It can contain large amounts of data and inherit existing functionalities in the relational database.
	\item MOLAP uses array-based multidimensional storage engines for multidimensional views of data, rather than in a relational database. So, it can perform complex calculations with good performance. Besides, it is usually used as an OLAP engine for a DW which is built on a multidimensional schema.
	\item HOLAP is a combination of both ROLAP and MOLAP. It uses both relational and multidimensional techniques to inherit the higher scalability of ROLAP and the faster computation of MOLAP.
	\end{enumerate}
		
	In our agricultural Big Data context, HOLAP is more suitable than ROLAP and MOLAP because:
	\begin{enumerate}
	\vspace{-1mm}
	\item ROLAP has quite slow performance. Each ROLAP report is an SQL query in the relational database that requires a significant execution time. In addition, ROLAP does not meet all the users' needs, especially complex queries.
	\item MOLAP requires that all calculations should be performed during the data cube construction. So, it handles only a limited amount of data and does not scale well. In addition, MOLAP is not capable of handling detailed data.
	\item HOLAP inherits relational technique of ROLAP to store large data volumes and detailed information. Additionally, HOLAP also inherits multidimensional techniques of MOLAP to perform complex calculations and has good performance.
	\end{enumerate}

	\vspace{-2mm}
	\subsection{The Proposed Architecture}
	\vspace{-2mm}
		
	Based on the analyis in Section 3, Hive is chosen for building our DW storage and it is combining with MongoDB to implement our Integrated Information module. This is for the following reasons:
	\begin{enumerate}
	\vspace{-2mm}
		\item Hive is based on Hadoop which is the most powerful tool of Big Data. Besides, HQL is similar to SQL which is familiar to the majority of users. Especially, Hive supports well high storage capacity, business intelligent and data science more than MongoDB and Cassandra. These features of Hive are useful to make an agricultural DW and apply data mining technologies.
		\item Hive does not have real-time performance so it needs to be combined with MongoDB or Cassandra to improve performance of our Integrated Information module.
		\item MongoDB is more suitable than Cassandra to complement Hive because: (1) MongoDB supports joint operation, full text search, ad-hoc query and second index which are helpful to interact with users. While Cassandra does not support these features; (2) MongoDB has the same master – slave structure with Hive that is easy to combine. While the structure of Cassandra is peer - to - peer; (3) Hive and MongoDB are more reliable and consistent. So the combination between Hive and MongoDB supports fully the CAP theorem while Hive and Cassandra are the same AP systems.
	\end{enumerate}

	\begin{figure}
	\vspace{-6mm}
	\begin{center}
             \includegraphics[width=1\textwidth, height=7.6cm]{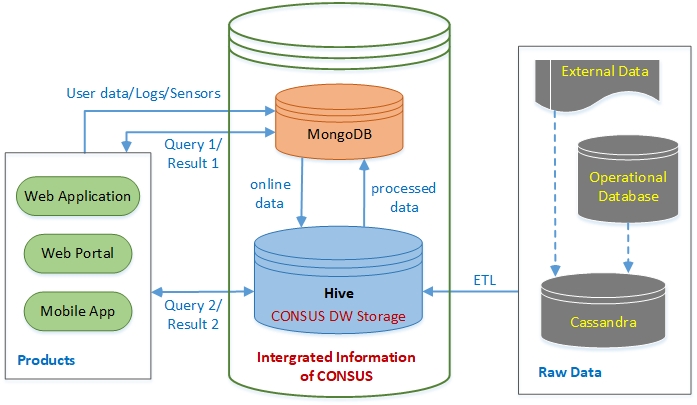}
	\end{center}
	\vspace{-6mm}
  		\caption{Our agricultural data warehouse architecture}
	\label{fig_DW_Storage}
	\vspace{-5mm}
	\end{figure}	
	
		Our DW architecture for agricultural Big Data is illustrated in Figure \ref{fig_DW_Storage} which contains three modules, namely Integrated Information, Products and Raw Data. The Integrated Information module includes two components being MongoDB component and Hive component.	Firstly, the MongoDB component will receive real-time data, such as user data, logs, sensor data or queries from Products module, such as web application, web portal or mobile app. Besides, some results which need to be obtained in real-time will be transferred from the MongoDB to Products. Second, the Hive component will store the online data from and send the processed data to the MongoDB module. Some kinds of queries having complex calculations will be sent directly to Hive. After that, Hive will send the results directly to the Products module.
		
		In Raw Data module, almost data in Operational Databases or External Data components is loaded into Cassandra component. It means that we use Cassandra to represent raw data storage. This improves the performance of ETL and helps us deploy our system on cloud or distributed systems better.
		
	\begin{figure}[h]
	\vspace{-5mm}
	\begin{center}
             \includegraphics[scale=0.59]{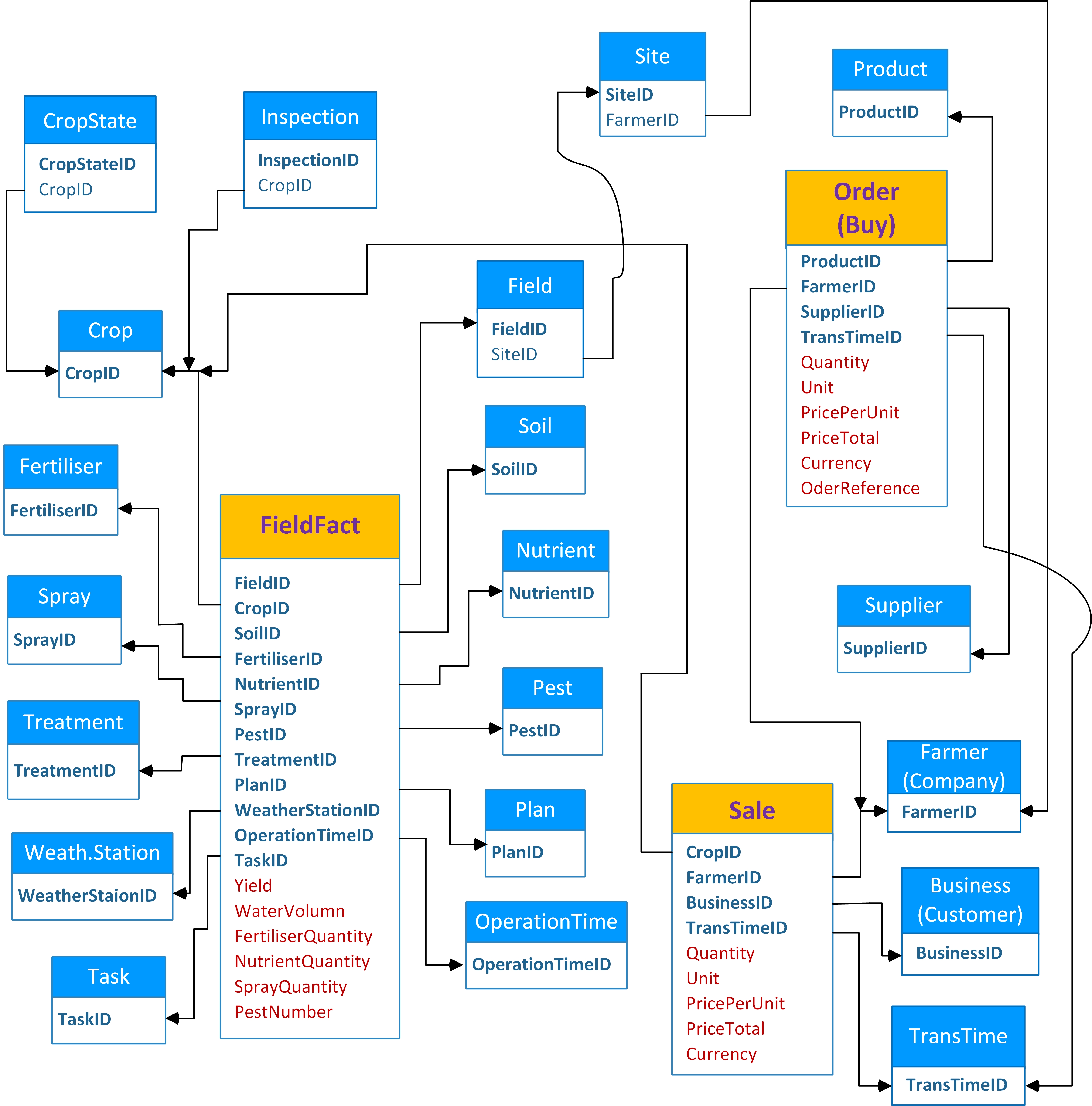}
	\end{center}
	\vspace{-4mm}
  		\caption{A part of our data warehouse schema for Precision Agriculture}
	\label{fig_Schema}
	\vspace{-3mm}
	\end{figure}	
	\vspace{-5mm}	
	
	\subsection{Our Schema}

	The DW uses schema to logically describe the entire datasets. A schema is a collection of objects, including tables, views, indexes, and synonyms which consist of some fact and dimension tables \cite{Oracle.2017}. The DW schema can be designed through the model of source data and the requirements of users. There are three kind of schemas, namely star, snowflake and constellation. With features of agricultural data, the agricultural DW schema needs to have more than one fact table and be flexible. So, the constellation schema, also known galaxy schema, is selected to design our DW schema.
	
	We developed a constellation schema for our agricultural DW and it is partially described in Figure \ref{fig_Schema}. It includes 3 fact tables and 19 dimension tables. The FieldFact fact table contains data about agricultural operations on fields. The Order and Sale fact tables contain data about farmers' trading operations. The FieldFact, Order and Sale facts have 12, 4 and 4 dimensions, and have 6, 6 and 5 measures, respectively. While, dimension tables contain details about each instance of an object involved in a crop yield. The main attributes of these dimension tables are described in the Table \ref{tab5}. The key dimension tables are connected to their fact table. However, there are some dimension tables connected to more than one fact table, such as Crop and Farmer. Besides, the CropState, Inspection and Site dimension tables are not connected to any fact table. The CropState and Inspection tables are used to support the Crop table. While, the Site table supports the Field table. 
				
		\begin{table}[!htbp]
		\vspace{-3mm}
		\caption{Descriptions of some dimension tables}
		\vspace{-5mm}
		\begin{center}
		\scriptsize
		\begin{tabular}{|c|c|m{96mm}|}
		\hline
		\textbf{No.} & \textbf {Dim. tables} & \textbf{Particular attributes}\\
		\hline
		1 & Business & BusinessID, Name,Address, Phone, Mobile, Email\\
		\hline
		2 & Crop & CropID, CropName, VarietyID, VarietyName, EstYield, SeasontSart, SeasonEnd, BbchScale, ScientificName, HarvestEquipment, EquipmentWeight\\
		\hline
		3 & CropState & CropStateID, CropID, StageScale, Height, MajorStage, MinStage, MaxStage, Diameter, MinHeight, MaxHeight, CropCoveragePercent \\
		\hline
		4 & Farmer & FarmerID, FarmerName, Address, Phone, Mobile, Email  \\
		\hline
		5 & Fertiliser & FertiliserID, Name, Unit, Status, Description, GroupName \\	
		\hline
		6 & Field & FieldID, FieldName, SiteID, Reference, Block, Area, AreaUnit, WorkingArea, WorkingAreaUnit, FieldGPS, Notes \\
		\hline
		7 & Inspection & InspectionID, CropID, Description, ProblemType, Severity, ProblemNotes, AreaValue, AreaUnit, Order, Date, Notes, GrowthStage \\
		\hline
		8 & Nutrient & NutrientID, NutrientName, Date, Quantity \\
		\hline
		9 & OperationTime & OperationTimeID, StartDate, EndDate, Season \\
		\hline
		10 & Pest & PestID, CommonName, ScientificName, PestType, Description, Density, MinStage, MaxStage, Coverage, CoverageUnit \\
		\hline
		11 & Plan & PlanID, PlanName, PlanNumber, RegistrationNo, ProductName, ProductRate, Date, WaterVolume \\
		\hline
		12 & Product & ProductID, ProductName, GroupName \\
		\hline
		13 & Site & SiteID, FarmerID, SiteName, Reference, Country, AddressName, AddressTown, PostalCode, GPS, Created, CreatedBy\\		
		\hline
		14 & Spray & SprayID, SprayProductName, ProductRate, AppliedArea, AppliedDate, WaterVolume, VolumeUnit, ConfirmDuration, ConfirmWindSPeed, ConfirmDirection, ConfirmTemperature, ConfirmHumidity, ActivityType \\
		\hline
		15 & Soil & SoilID, PH, Phosphorus, Potassium, Magnesium, Calcium, CEC, Silt, Clay, Sand, TextureLabel, TestDate \\
		\hline
		16 & Supplier & SupplierID, SupplierName, SupplierContactName, Address, ContactPhone, ContactMobile, ContactEmail\\
		\hline
		17 & Task & TaskID, TaskDesc, TaskStatus, TaskDate, TaskInterval, CompletedDate, AppCode \\
		\hline
		18 & Treatment & TreatmentID, TreatmentName, FormType, LotCode, Rate, ApplCode, LevlNo, Type, Description, ApplDesc, TreatmentComment \\
		\hline
		19 & WeatherStation & WeatherStationID, StationName, MeasureDate, AirTemperature, SoilTemperature, StationReadingBatch \\
		\hline
		\end{tabular}
		\label{tab5}
		\end{center}
		\vspace{-8mm}
		\end{table}

\vspace{-2mm}
\section{Experiments}
\vspace{-2mm}

	Through the proposed architecture in Section 4.2, our DW inherited many advantages from Hive, MongoDB and Cassandra presented in Section 3, such as high performance, high storage, large scale analytic and security. In the scope of this paper, we evaluated our DW schema and data analysis capacity on real agricultural datasets through complex queries. In addition, the time performance of our agricultural DW storage was also evaluated and compared to MySQL on many particular built queries belonging to different query groups.
		
	\vspace{-2mm}	
	\subsection{Data Analyzing Demo}
		\vspace{-2mm}

		\begin{wrapfigure}{r}{0.52\textwidth}
		\vspace{-2mm}
		\begin{center}
             \includegraphics[scale=0.6]{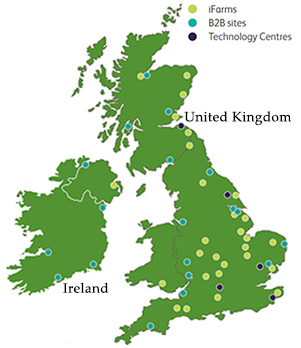}
		\end{center}
		\vspace{-5mm}
  		\caption{Data in UK and Ireland \cite{Origin.2018}}
		\label{fig_UKIreland}
		\vspace{-3mm}
		\end{wrapfigure}
		
\begin{figure}[H]
		\vspace{-2mm}
		\begin{center}
             \includegraphics[scale=0.69]{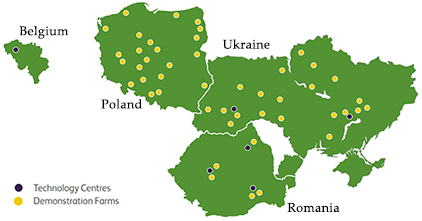}
		\end{center}
		\vspace{-4mm}
  		\caption{Data in Continental Europe \cite{Origin.2018}}
  		\vspace{-5mm}
		\label{fig_continentaleurope}
		\vspace{-2mm}
		\end{figure}

	The input data for the DW was  primarily obtained from an agronomy company which supplies data from its operational systems, research results and field trials. Specially, we are supplied real agricultural data in iFarms, B2B sites, technology centres and demonstration farms. Their specific positions in several European countries are presented in Figures \ref{fig_UKIreland} and \ref{fig_continentaleurope} \cite{Origin.2018}. There is a total of 29 datasets. On average, each dataset contains 18 tables and is about 1.4 GB in size. The source datasets are loaded on our CONSUS DW Storage based on the schema described in Section 4.3 through an ETL tool. From the DW storage, we can extract and analyze useful information through tasks using complex HQL queries or data mining algorithms. These tasks could not be executed if the separate 29 datasets have not been integrated into our DW storage.

\begin{figure}[H]
	\vspace{-5mm}
	\begin{center}
             \includegraphics[scale=0.49]{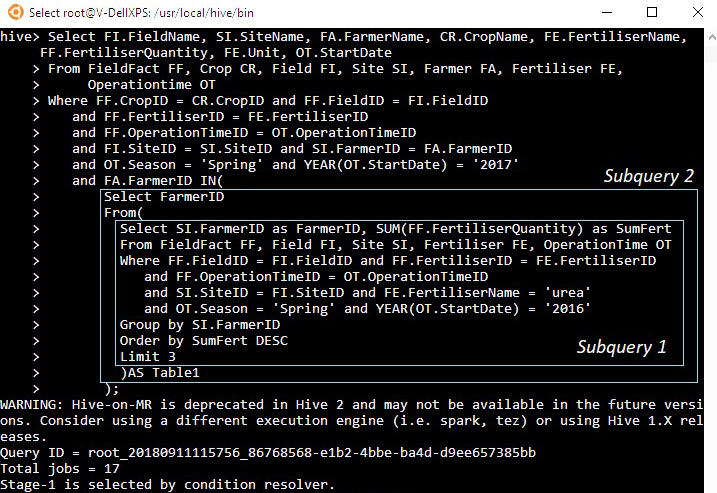}
	\end{center}
	\vspace{-5mm}
  	\caption{A screenshort of executing the query example in our Hive}
  	\vspace{-5mm}
	\label{fig_demo}
	\end{figure}

		
	An example for a complex request: \textit{''List crops, fertilisers, corresponding fertiliser quantities in spring, 2017 in every field and site of 3 farmers (crop companies) who used the large amount of Urea in spring, 2016''}. In our schema, this query can be executed by a HQL/SQL query as shown in Figure \ref{fig_demo}. To execute this request, the query needs to exploit data in the FieldFact fact table and the six dimension tables, namely Crop, Field, Site, Farmer, Fertiliser and OperationTime. The query consists of two subqueries. It returned \textit{3 farmers (crop companies) that used the largest amount of Urea in spring, 2016}.
	
	\vspace{-5mm}
	\subsection{Performance Analysis}
	\vspace{-1mm}
	
		The performance analysis was implemented using MySQL 5.7.22, JDK 1.8.0\_171, Hadoop 2.6.5 and Hive 2.3.3 which run on Bash on Ubuntu 16.04.2 on Windows 10. All experiments were run on a laptop with an Intel Core i7 CPU (2.40 GHz) and 16 GB memory. We only evaluate reading performance of our DW storage because a DW is used for reporting and data analysis. The database of our storage is duplicated into MySQL to compare performance. By combining popular HQL/SQL commands, namely Where, Group by,  Having, Left (right) Join, Union and Order by, we create 10 groups for testing. Every group has 5 queries and uses one, two or more commands (see Table \ref{tab_queries}). Besides, every query also uses operations, such as And, Or, $\ge$, Like, Max, Sum and Count, to combine with the commands.
	
	All queries were executed three times and we took the average value of the these executions. The different times in runtime between MySQL and our storage of query $q_i$ is calculated as $Times_{q_i} = RT^{mysql}_{q_i}/RT^{ours}_{q_i}$. Where, $RT^{mysql}_{q_i}$ and $RT^{ours}_{q_i}$ are respectively average runtimes of query $q_i$ on MySQL and our storage. Besides, with each group $G_i$, the different times in runtime between MySQL and our storage $Times_{G_i} = RT^{mysql}_{G_i}/RT^{ours}_{G_i}$. Where, $RT_{G_i} = Average(RT_{q_i})$
is average runtime of group $G_i$ on MySQL or our storage.

		\begin{table}
		\vspace{-3mm}
		\caption{Command combinations of queries}
		\begin{center}
		\scriptsize
		\begin{tabular}{|c|c|c|c|c|c|c|c|}
		\hline
		\textbf{Group} & \textbf {Queries} & \textbf{Where} & \textbf{Group by}  & \textbf{Having} & \textbf{Left (right) Joint} & \textbf{Union} & \textbf{Order by}\\
		\hline
		1 & 1 - 5 & x & & & & & \\
		\hline
		2 & 6 - 10 & x & x & & & & \\
		\hline
		3 & 11 - 15 & x & & & x & & \\
		\hline
		4 & 16 - 20 & x & & & & x & \\
		\hline
		5 & 21 - 25 & x & & & & & x\\
		\hline
		6 & 26 - 30 & x & & & x & & x\\
		\hline
		7 & 31 - 35 & x & x & x & & & \\
		\hline
		8 & 36 - 40 & x & x & x & & & x \\
		\hline
		9 & 41 - 45 & x & x & x & x & & x \\
		\hline
		10 & 45 - 50 & x & x & x & & x & x \\
		\hline

		\end{tabular}
		\end{center}
		\vspace{-3mm}
		\label{tab_queries}
		\vspace{-5mm}
		\end{table}
		
		
		\begin{figure}[H]
		\vspace{-6mm}
		\begin{center}
		\begin{tikzpicture}
		\pgfplotsset{height=72mm, width= 116mm, xlabel near ticks, ylabel near ticks}
		\begin{axis}[
		font=\footnotesize,
		enlargelimits=0.15,
		legend style={draw=none, fill=none, font=\footnotesize, cells={anchor=west} ,  legend pos=north east,  xshift= -42mm,yshift= 1mm },  xlabel={Queries ($q_i$)},  ylabel={Different times ($Times_{q_i}$)},
		]
		
		\addplot+[ycomb] plot coordinates {(1, 6.3) (2, 14.8) (3, 1.9) (4, 13.5) (5, 33) };
		\addplot+[ycomb] plot coordinates {(6, 29.1) (7, 3.2) (8, 3.5) (9, 2.3) (10, 2.1) };
		\addplot+[ycomb] plot coordinates {(11, 25.3) (12, 0.5) (13, 0.8) (14, 26) (15, 14.5) };
		\addplot+[ycomb] plot coordinates {(16, 3.4) (17, 1.3) (18, 0.8) (19, 4.8) (20, 1.6) };
		\addplot+[ycomb] plot coordinates {(21, 4.5) (22, 2.2) (23, 1.6) (24, 2.8) (25, 1) };
		\addplot+[ycomb] plot coordinates {(26, 2.3) (27, 5.4) (28, 9.7) (29, 20.3) (30, 2.8) };
		\addplot+[ycomb] plot coordinates {(31, 2.7) (32, 2.6) (33, 37.6) (34, 3.4) (35, 18.4) };
		\addplot+[ycomb] plot coordinates {(36, 35.1) (37, 4.1) (38, 23.5) (39, 5.3) (40, 1.6) };
		\addplot+[ycomb] plot coordinates {(41, 12.6) (42, 2.2) (43, 2.1) (44, 19.6) (45, 11.1) };
		\addplot+[ycomb] plot coordinates {(46, 1.1) (47, 2.4) (48, 1.2) (49, 1.9) (50, 1.4) };
		
		\addplot[purple,sharp plot,update limits=false]
		coordinates {(-4,1) (53,1)}
		node[above] at (axis cs:-4.2,0.8) {1};
		
		\legend{Group 1, Group 2, Group 3, Group 4, Group 5, Group 6, Group 7, Group 8, Group 9, Group 10}
		\end{axis}
		\end{tikzpicture}
		
		\end{center}
		\vspace{-5mm}
		\caption{Different times between MySQL and our storage in runtime every Query}
		\label{fig:times50queries}
		\vspace{-5mm}
		\end{figure}
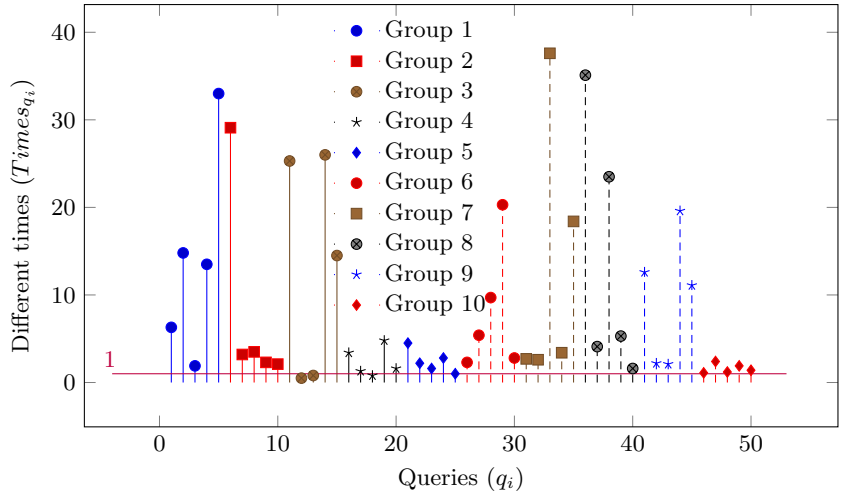
		
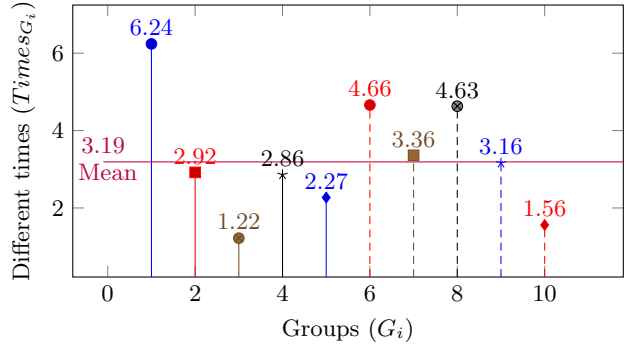
\begin{figure}[H]
		\vspace{-5mm}
		\begin{center}
		\begin{tikzpicture}
		\pgfplotsset{height=52mm, width= 89mm, xlabel near ticks, ylabel near ticks}
		\begin{axis}[
		font=\footnotesize,
		enlargelimits=0.2,
		legend style={draw=none, fill=none, font=\footnotesize, cells={anchor=west} ,  legend pos=north east,  xshift= 25mm,yshift= 2mm }, 
		xlabel={Groups ($G_i$)},  ylabel={Different times ($Times_{G_i}$)},
		nodes near coords,
		nodes near coords align={vertical},
		]
		
		\addplot+[ycomb] plot coordinates {(1, 6.24)};
		\addplot+[ycomb] plot coordinates {(2, 2.92)};
		\addplot+[ycomb] plot coordinates {(3, 1.22)};
		\addplot+[ycomb] plot coordinates {(4, 2.86)};
		\addplot+[ycomb] plot coordinates {(5, 2.27)};
		\addplot+[ycomb] plot coordinates {(6, 4.66)};
		\addplot+[ycomb] plot coordinates {(7, 3.36)};
		\addplot+[ycomb] plot coordinates {(8, 4.63)};
		\addplot+[ycomb] plot coordinates {(9, 3.16)};
		\addplot+[ycomb] plot coordinates {(10, 1.56)};
		
		\addplot[purple,sharp plot,update limits=false]
		coordinates {(-0.1,3.19) (12,3.19)}
		node[above] at (axis cs:0, 2.5) {Mean};
		
		\end{axis}
		\end{tikzpicture}
		
		\end{center}
		\vspace{-5mm}
		\caption{Different times between MySQL and our storage in runtime of every group}
		\label{fig:times10groups}
		\vspace{-5mm}
		\end{figure}

		
		Figure \ref{fig:times50queries} and Figure \ref{fig:times10groups} describe different times in runtime in every query belongs to 10 groups and every group. Unsurprisingly, although running on one computer, but with large data volume, our storage is faster than MySQL at 46/50 queries and all 10 query groups. MySQL is faster than our storage at 3 queries $12^{th}$, $13^{th}$ and $18^{th}$ belonging to groups $3^{rd}$ and $4^{th}$. Two databases are same at the query $25^{th}$ belonging to group $5^{th}$. Comparing to MySQL, our storage is more than at most (6.24 times) at group $1^{st}$ which uses only \textit{Where} command, and at least (1.22 times) at group $3^{rd}$ which uses \textit{Where} and \textit{Joint} commands. 
			
\begin{figure}[H]
		\vspace{-5mm}
		\begin{center}
		\begin{tikzpicture}
		\pgfplotsset{height=59mm, width= 116mm, xlabel near ticks, ylabel near ticks}
		\begin{axis}[
		    font=\footnotesize,
		    ybar,
		    bar width=2.7pt,
		    enlargelimits=0.15,
		    legend style={draw=none, fill=none, font=\footnotesize, cells={anchor=west} ,  legend pos=north east,  xshift= 1mm,yshift= 0mm}, 
		    xlabel={Groups},
		    ylabel={Average runtimes (seconds)},
		    symbolic x coords={1, 2, 3, 4, 5, 6, 7, 8, 9, 10, Mean },
		    xtick=data,
		    nodes near coords,
		    nodes near coords align={vertical},
		    ]
			
		\addplot coordinates {(1, 1081.5) (2, 599.7) (3, 111.7) (4, 790.4) (5, 776.6) (6, 1109.2) (7, 483) (8, 1057.3) (9, 297.9) (10, 571.1) (Mean, 687.8)};
		\addplot coordinates {(1, 173.4) (2, 205.2) (3, 91.2) (4, 276.4) (5, 342.8) (6, 238) (7, 143.7) (8, 228.3) (9, 94.2) (10, 366.4) (Mean, 216.1)};
				
		\legend{MySQL, Our storage}
		\end{axis}
		\end{tikzpicture}
		\end{center}
		\vspace{-5mm}
		\caption{Average Runtimes of MySQL and our storage in every Groups}
		\label{fig:runtime_groups}
		\vspace{-6mm}
		\end{figure}
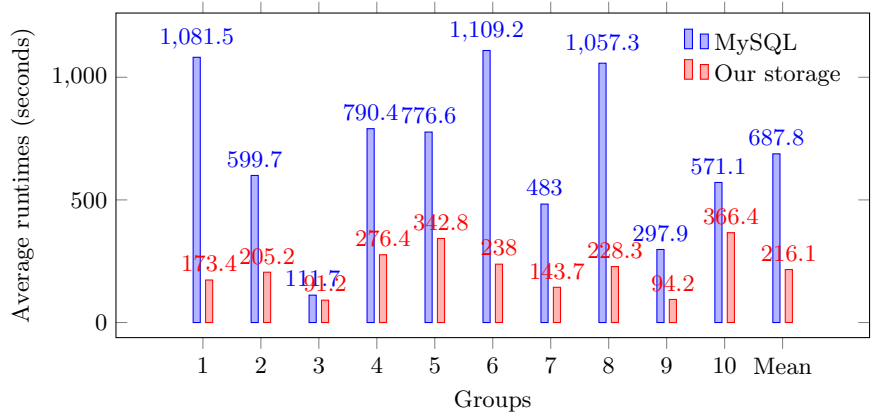

		Figure \ref{fig:runtime_groups} presents the average runtime of the 10 query groups on MySQL and our storage. Mean, the run time of a reading query on MySQL and our storage is 687.8 seconds and 216.1 seconds, respectively. It means that our storage is faster 3.19 times. In the future, by deploying our storage solution on cloud or distributed systems, we believe that the performance will be even much better than MySQL.
		
\vspace{-3mm}		
\section{Conclusion}
\vspace{-2mm}
In this paper, we compared and analyzed some existing popular open source DWs in the context of agricultural Big Data. We designed and implemented the agricultural DW by combining Hive, MongoDB and Cassandra DWs to exploit their advantages and overcome their limitations. Our DW includes necessary modules to deal with large scale and efficient analytics for agricultural Big Data. Additionally, the presented schema herein was optimised for the real agricultural datasets that were made available to us. The schema been designed as a constellation so it is flexible to adapt to other agricultural datasets and quality criteria of agricultural Big Data. Moreover, using the short demo, we outlined a complex HQL query that enabled knowledge extraction from our DW to optimize of agricultural operations. Finally, through particular reading queries using popular HQL/SQL commands, our DW storage outperforms MySQL by far.

In the future works, we shall pursue the deployment of our agricultural DW on a cloud system and implement more functionalities to exploit this DW. The future developments will include: (1) Sophisticated data mining techniques to determine crop data characteristics and combine with expected outputs to extract useful knowledge; (2) Predictive models based on machine learning or graph algorithms \cite{Helmer.2015}; (3) An intelligent interface for data access; (4) Combination with the high-performance knowledge map framework \cite{Ngo.2011}.

\vspace{-2mm}
\section*{Acknowledgment}
\vspace{-3mm}

This research is part of the CONSUS research programme which is funded under the SFI Strategic Partnerships Programme (16/SPP/3296) and is co-funded by Origin Enterprises Plc.

%

%

\end{document}